\documentclass[prd,aps,nofootinbib,onecolumn]{revtex4}
\usepackage{amsmath}
\usepackage{graphicx}
\usepackage{amsmath}
\usepackage{amssymb}
\usepackage{latexsym}
\usepackage{color}

\newcommand{\GB}{\mathcal{G}}
\newcommand{\de}{\mathrm{d}}

\begin{document}

\title{Cosmological perturbation in $f(R,\GB)$ theories with a perfect fluid}

\author{Antonio De Felice} \email{defelice@rs.kagu.tus.ac.jp}
\affiliation{Department of Physics, Faculty of Science, Tokyo University of Science, 1-3 Kagurazaka, Shinjuku-ku, Tokyo, 162-8601, Japan}

\author{Jean-Marc G\'erard}
\email{jean-marc.gerard@uclouvain.be}
\affiliation{Centre for Particle Physics and Phenomenology (CP3),
  Universit\'e catholique de Louvain, Chemin du Cyclotron 2, B-1348
  Louvain-la-Neuve, Belgium}

\author{Teruaki Suyama}
\email{teruaki.suyama@uclouvain.be}
\affiliation{Research Center for the Early Universe, Graduate School of Science, The University of Tokyo, Tokyo 113-0033, Japan}
\affiliation{Centre for Particle Physics and Phenomenology (CP3), Universit\'e catholique de Louvain, Chemin du Cyclotron 2, B-1348 Louvain-la-Neuve, Belgium}

\date{\today}

\begin{abstract}
In order to classify modified gravity models according to their physical properties,
we analyze the cosmological linear perturbations for $f(R,\GB)$ theories (R being the Ricci scalar and $\GB$, the Gauss-Bonnet term) with a minimally coupled perfect fluid. For the scalar type perturbations, we identify in general six degrees of freedom. We find that two of these physical modes obey the same dispersion relation as the one for a non-relativistic de Broglie wave. This means that spacetime is either highly unstable or its fluctuations undergo a scale-dependent super-luminal propagation. Two other modes correspond to the degrees of freedom of the perfect fluid, and propagate with the sound speed of such a fluid. The remaining two modes correspond to the entropy and temperature perturbations of the perfect fluid, and completely decouple from the other modes for a barotropic equation of state.  We then provide a concise condition on $f(R,\GB)$ theories, that both $f(R)$ and $R+f(\GB)$ do fulfill, to avoid the de Broglie type dispersion relation. For the vector type perturbation, we find that the perturbations decay in time. For the tensor type perturbation, the perturbations can be either super-luminal or sub-luminal, depending on the model. No-ghost conditions are also obtained for each type of perturbation.
\end{abstract}

\maketitle

\section{Introduction}

 Many modified gravity theories have been introduced in order to explain the acceleration of the universe \cite{review}. In particular, the first working model of inflation---the Starobinsky model \cite{Starob}---made use of such modifications that should somehow appear in any effective Lagrangian for gravity. These modifications are usually considered to be negligible at low energies. Yet, we do not know much about the reference scale for gravity. Indeed, there might be two fundamental scales defined by the Planck mass and the cosmological constant. At this level of ignorance, people have considered models of dark energy that change only the gravity sector. It amounts to substitute general functions of the Riemann tensor for the Einstein-Hilbert $R$ term in the action \cite{fRori,fRviable,fRmatter,PQR}.

If we consider such a departure from the single $R$ term to be responsible for dark energy, it should become important only at late times. But if its contribution to the action becomes stronger and stronger as time passes by, the full Lagrangian has to be considered as the fundamental one with its own particle content. This implies for example that $f(R)$ theories indeed possess an extra scalar propagating field \cite{Hanlon}.

Along these lines attempts based on the other non-zero Lovelock scalar, the Gauss-Bonnet combination $\GB$, also appeared. In the so called Gauss-Bonnet theories, the Lagrangian for gravity reads ${\cal L}=R+f(\GB)$ \cite{GaussBN}. Recently a paper \cite{TsujiInst} showed the presence of classical matter instabilities in such models, ruling them out as a sensible theory for gravity and dark energy. Yet, in addition to $f(R)$ and $R+f(\GB)$ theories, one still has the freedom of introducing the most general Lovelock modification of gravity, that is ${\cal L}=f(R,\GB)$. Naively, one may expect that the stable perturbation behavior of $f(R)$ can overcome the unstable behavior of $f(\GB)$. However, the story is not so simple. For the vacuum case (i.e., absence of matter), it was shown in \cite{DeFelice09} that if the theory gives a non-vanishing $f_{RR} f_{\GB \GB}-f_{R \GB}^2$, then the nature of scalar perturbations qualitatively differs from the one for a class of theories to which $f(R)$ and $R+f(\GB)$ belong. The study was extended to a system where a single scalar field is (not necessarily minimally) coupled to gravity and similar results were obtained \cite{DeSuScal}. 

It is necessary to develop theoretical and experimental tools to disentangle all these alternative theories of gravity. 
We believe they should be classified according to some physical properties. In this paper we argue that
the high-$k$ modes behavior, i.e. their scale dependent speed of propagation, in the presence of a perfect fluid (not only
in vacuum as in \cite{DeFelice09}) can provide such a key physical feature according to which the various theories of gravity can be classified.

The scope of this paper is to develop the cosmological linear perturbation theory \cite{cosmoreview} for the general $f(R,\GB)$ theories in the presence of a perfect fluid. In particular we will find conditions in order to remove ghosts \cite{ghost1}---by imposing the kinetic operator for the fields to be positive definite---and Laplacian instabilities sourced by a negative squared speed for the propagating fields. Clearly, this study is relevant since our universe contains matter components which are well described by a perfect fluid. Therefore, the results obtained in this paper not only reveal the theoretical structure of the model but are also useful for the purpose of constraining modified gravity theories from observations of the cosmic structures.

In section \ref{sec:action} we introduce the action and write down the background equations of motion. Linear perturbation theory for the scalars are discussed in Section \ref{sec:scalar}. Vector and tensor modes are discussed in Sections \ref{sec:vector} and \ref{sec:tensor} respectively. Our conclusions are given in Section \ref{sec:concl}.

\section{The action}\label{sec:action}
The action we study is given by
\begin{equation}
  \label{eq:act1}
  S=\int d^4x\sqrt{-g}\left[\frac{f(R,\GB)}{16\pi G}+ p(\mu,s)\right]\, ,
\end{equation}
where $G$ is the Newton constant and $p$ is the pressure of a perfect fluid characterized by a chemical potential $(\mu)$ and an entropy per particle $(s)$. 
The four-velocity of the perfect fluid is given by potentials,
\begin{equation}
u_\nu=\frac{1}{\mu} (\partial_\nu \ell+\theta \partial_\nu s +A \partial_\nu B), \label{four-velocity}
\end{equation}
where $\ell,\theta,A$ and $B$ are all scalar quantities.
The action (\ref{eq:act1}) for a perfect fluid has been introduced in \cite{Schutz}, and its application to the cosmological perturbations for a single perfect fluid within General Relativity(GR) has been developed in a recent paper \cite{DeGeSu}. The fundamental fields that should be variated to derive the equations of motion are $g_{\mu \nu},\ell,\theta,A,B$ and $s$ ($\mu$ is written in terms of these fundamental fields by using the normalization condition $u^\mu u_\mu=-1$). The variation with respect to $g_{\mu \nu}$ yields the gravitational field equations:
\begin{equation}
R_{\mu \nu}-\tfrac{1}{2} g_{\mu \nu}R-\Sigma_{\mu \nu}=8\pi G\,T_{\mu\nu}, \label{eom}
\end{equation}
where $\Sigma_{\mu \nu}$ is the effective energy momentum tensor defined by
\begin{align} 
\Sigma_{\mu \nu}&=\nabla_\mu \nabla_\nu F-g_{\mu \nu} \Box F+2R \nabla_\mu \nabla_\nu \xi-2g_{\mu \nu} R \Box \xi-4R_\mu^{~\lambda} \nabla_\lambda \nabla_\nu \xi-4R_\nu^{~\lambda} \nabla_\lambda \nabla_\mu \xi 
+4R_{\mu \nu} \Box \xi \nonumber \\
&\qquad+4 g_{\mu \nu} R^{\alpha \beta} \nabla_\alpha \nabla_\beta \xi+4R_{\mu \alpha \beta \nu} \nabla^\alpha \nabla^\beta \xi-\tfrac{1}{2}\,g_{\mu \nu} V +(1-F)\,\bigl(R_{\mu \nu}-\tfrac{1}{2} g_{\mu \nu}R\bigr). \label{effective-energy-momentum}
\end{align}
In Eq.~(\ref{effective-energy-momentum}), $V\equiv FR+\xi\,\GB-f(R,\GB)$, and 
\begin{equation}
  \label{eq:def1}
  F\equiv\frac{\partial f}{\partial R}\,,\qquad\xi\equiv \frac{\partial f}{\partial\GB}.
\end{equation}
Meanwhile, the variations with respect to $\ell,\theta,s,A$ and $B$, with
a help of the first law of thermodynamics $\de p=n \de\mu-nT \de s$ ($n$ being the number density and $T$, the temperature), yield the following equations of motion:
\begin{align}
&\partial_\alpha \left( n \sqrt{-g} u^\alpha \right)=0,~~~~~u^\alpha \partial_\alpha s=0,~~~~~u^\alpha \partial_\alpha \theta -T=0, ~~~~~u^\alpha \partial_\alpha A=0,~~~~~u^\alpha \partial_\alpha B=0. \label{eqfluid}
\end{align}
The first equation represents the conservation of the particle number and
the second one, the conservation of the entropy.
From Eqs.~(\ref{eqfluid}) and the relation $\rho=n\mu-p$, 
we can derive the standard energy-momentum conservation \cite{Schutz}:
\begin{equation}
\nabla_\mu T^\mu_{~\nu}=\nabla_\mu [ (\rho+p)u^\mu u_\nu +p\, \delta^\mu_{~\nu}]=0.
\end{equation}

Before going to the linear perturbation theory, let us first derive the evolution equations
for the Friedmann-Lema\^\i tre-Robertson-Walker(FLRW) flat universe whose metric is given by
\begin{equation}
ds^2=-dt^2+a^2(t) dx^i dx_i,
\end{equation}
with $u^\mu=(1,0,0,0)$, the four-velocity and $H={\dot a}/a$, the Hubble parameter.
In a FLRW universe, all the physical variables for a perfect fluid depend only on the cosmic time $t$:
\begin{equation}
\rho=\rho(t),~~~~p=p(t),~~~~n=n(t),~~~~T=T(t),~~~~\mu=\mu (t),~~~~s=s(t).
\end{equation}
On the other hand, the velocity potentials can depend on the spatial coordinates: 
\begin{equation}
A=A({\vec x}),~~~~B=B({\vec x}),~~~~\theta=\int^t \de t' ~T(t')+{\tilde \theta} ({\vec x}),
\end{equation}
where $A,B$ and ${\tilde \theta}$ are arbitrary functions leading to equivalent physical backgrounds consistent
with homogeneity and isotropy.
We will take advantage of this freedom to simplify our perturbation studies.

In a FLRW flat universe, the gravitational field equations (\ref{eom}) reduce to 
\begin{align}
\label{eq:defV}
3H^2&=\frac{1}{F} \left[\tfrac12\, V-3H{\dot F}-12H^3 {\dot \xi}+8\pi G\,\rho\right],\\
\ddot F&=4\, \dot\xi  H ^{3}-4\, \ddot\xi  H ^{2} + \dot F H  -8\, \dot\xi H  \dot H -8\,\pi \,G(\rho+p)-2\,F \dot H
\label{eqzFB}
\end{align}
and allow us to remove respectively $V$ and $\ddot F$ in all the remaining calculations. 
The continuity equation $\dot\rho+3H(\rho+p)=0$ is a direct consequence of the previous two equations.
Indeed, 
\begin{align}
R&=6(2H^2+{\dot H}), \\
\GB&=24 H^2 (H^2+{\dot H})=24 H^2 \frac{\ddot a}{a},
\end{align}
with $\sqrt{-g} \GB=8\, \de({\dot a}^3)/\de t$ a total derivative with respect to time, as it should be.

\section{Scalar modes}\label{sec:scalar}
For the scalar gravity perturbations, we expand the metric as
\begin{equation}
ds^2=-(1+2\alpha)\,dt^2-2 a(t) \partial_i \beta\, dt\, dx^i+a^2(t)\, (\delta_{ij}+2\phi \delta_{ij}+2\partial_i \partial_j \gamma )\, dx^i\, dx^j.
\end{equation}
For the scalar matter perturbations, we introduce $\delta\ell$, $\delta \theta$, $\delta s$, $\delta A$ and $\delta B$ and 
consider the simplest solution for $A,B$ and ${\tilde \theta}$, namely
\begin{equation}
A=B={\tilde \theta}=0.
\end{equation}
With this particular choice, the scalar velocity perturbation defined by $\delta u_i=\mu^{-1}\partial_i v$ 
is related to the potentials introduced in Eq.~(\ref{four-velocity}) as $v=\delta \ell+\theta(t) \,\delta s$. Therefore we also have that $\delta T^0{}_i=n\,\partial_i v$.

To study the perturbation dynamics, we have to expand the action at second order in the fields.
As it is well known, the perturbation variables are not necessarily dynamical fields and we can reduce the second order action 
by eliminating the auxiliary ones. After integrations by parts, 
one is left with a Lagrangian that contains only gauge invariant fields and the action for the perturbations can be written as follows
\begin{alignat}{4}
S^{(2)}&=\tfrac12\,\int \de t \de^3 x\, a^3\, \bigl[ && A_{ab}{\dot V_a} {\dot V_b}
-a^{-2}B\epsilon_{ab} {\vec \nabla} {V_a} \cdot {\vec \nabla} \dot V_b
-a^{-4} D_{11} (\vec\nabla^2 V_1)^2
-a^{-2}E_{ab} {\vec \nabla} V_a \cdot {\vec \nabla} V_b \notag \\
&&& -C\epsilon_{ab} {\dot V_a} V_b-M_{ab}^2 V_a V_b-a^{-2}E_{13}\vec\nabla\delta s\cdot\vec\nabla V_1 
-C_{13}\dot V_1\delta s-C_{23}\dot V_2\delta s\notag\\
&&&-2\,m_{23}\, V_2\,\delta s-K\delta s^2+n\,(\dot{\delta\theta}_M\delta s-\dot{\delta s}\delta\theta_M+\dot{\delta A}\delta B-\dot{\delta B}\delta A)
\bigr], \label{secondaction}
\end{alignat}
with $\epsilon_{ab}=-\epsilon_{ba}$ and $\epsilon_{12}=1$. In Eq.~(\ref{secondaction}), we named $V_1\equiv\phi_M$, $V_2\equiv v_M$, and $\delta\theta_M$ with 
\begin{align}
\phi_M&=\phi-\frac{H\,(\delta F+4H^2\delta\xi)}{\dot F+4H^2\dot\xi}\,, \\
v_M&=v+\frac{\rho+p}{n}\,\frac{\delta F+4H^2\delta\xi}{\dot F+4H^2\dot\xi}\,,\\
\delta\theta_M&=\delta\theta-\frac{T\,(\delta F+4H^2\delta\xi)}{\dot F+4H^2\dot\xi}\, .
\end{align}

The reduced action (\ref{secondaction}) whose coefficients are given in appendix \ref{coeffo}
contains six fields : $V_1,V_2,\delta s, \delta \theta_M, \delta A$ and $\delta B$.
Among them, $\delta A$ and $\delta B$ completely decouple from the other fields.
We can also see that $\delta A$ and $\delta B$ do not contribute to the four-velocity (see Eq.\ (\ref{four-velocity})) since $A=B=0$ at the background level. 
Therefore, they never affect physical quantities and can be neglected in the following.

While $V_1$ and $V_2$ have terms that are quadratic in their time derivatives, $\delta s$ and $\delta \theta_M$ have only terms that are linear in their time derivatives. 
By taking a variation with respect to $\delta \theta_M$, we find that $\delta s$ is a constant of motion, $\delta s=\delta s(\vec x)$.
This fact implies that in the equations of motion for the remaining fields, it should be considered as a source. The variation for $\delta s$ gives the dynamics of $\delta\theta_M$ as follows
\begin{equation}
\label{eq:dthe}
2 n\dot{\delta \theta}_M-2K\delta s-2m_{23}\,v_{M}+\frac{E_{13}}{a^2}\,\vec\nabla^2 \phi_M-C_{13}\dot\phi_M-C_{23}\dot v_M=0\, .
\end{equation}
The equation of motion for $V_a$ becomes
\begin{align}
\frac1{a^{3}}\frac{\de}{\de t}{\left( a^3 A_{ab} {\dot V_b} \right)}&-\frac{B \epsilon_{ab}}{a^2} \vec\nabla^2 {\dot V_b}
-C \epsilon_{ab} {\dot V_b}+\frac{D_{11}}{a^4} \delta_{a1} \vec\nabla^4 V_1
-\frac1{a^2}\left[ E_{ab}+\tfrac12(\dot B+HB)\epsilon_{ab} \right] \vec\nabla^2 V_b\notag\\
&+\left[ M_{ab}^2-\tfrac12(\dot C+3HC)\epsilon_{ab} \right] V_b
-\frac{E_{13}}{2a^2} \delta_{a1} \vec\nabla^2 \delta s 
-\frac{\delta s}{2a^3}\frac{\de}{\de t}(a^3 C_{a3}) +m_{23}\,\delta_{a2}\,\delta s=0. \label{eqforVa}
\end{align}
Since there is no coupling between $\delta\theta_M$ and $\phi_M$ or $v_M$, we can first solve the equations of motion for the two latter fields. Substituting the obtained solution into Eq.\ (\ref{eq:dthe}), we can eventually determine the time evolution of $\delta \theta_M$. For a perfect fluid of barotropic equation of state $p=p(\rho)$, we find that the entropy perturbation $\delta s$ decouples from the other modes in the action (\ref{secondaction}) (see appendix \ref{fluido}, case I) such that $V_1$ and $V_2$ form a closed set of evolution equations. For $p=p(T)$ and $f_{RR}f_{\GB\GB}-f_{R\GB}^2=0$, then we have instead $K=0$, and there is no quadratic term for the entropy perturbation in the action (\ref{secondaction}) (see appendix \ref{fluido}, case II).

\subsection{Dispersion relation}
Dispersion relations are necessary in order to understand the Laplacian instabilities for the perturbations. 
So let us derive them for the short wavelength modes for which the time evolution of the background universe can be neglected.
We find that both $B$ and $D_{11}$ factors are proportional to the combination (see appendix \ref{coeffo}):
\begin{equation}
  \label{eq:def2}
  \Delta\equiv f_{RR}f_{\GB\GB}-f_{R\GB}^2=F_R \xi_\GB-F_\GB \xi_R.
\end{equation}
The qualitative behavior of the perturbations crucially depends on this quantity.
Indeed, let us first assume that the modified gravity theory satisfies $\Delta \neq 0$.
In this case, after performing a Fourier transformation, Eq.\ (\ref{eqforVa}) can be approximated, for large $k$'s, as
\begin{equation}
A_{ab} {\ddot V_b}+\frac{B \epsilon_{ab}k^2}{a^2} {\dot V_b}+\frac{D_{11}k^4}{a^4} \delta_{a1} V_1
+\left[ E_{ab}+\tfrac12\,(\dot B+HB)\, \epsilon_{ab} \right] \frac{k^2}{a^2}V_b=0. \label{appeqforVa}
\end{equation}
From this equation, we can derive the dispersion relations for four modes. 
The dispersion relation for the first two modes is given by
\begin{equation}
  \label{eq:cs2k2}
  \omega_1^2\approx\frac{B^2+A_{22}D_{11}}{A_{11}A_{22}-A_{12}^2}\,\frac{k^4}{a^4}. 
\end{equation}
This is exactly the same relation as the one for the non-relativistic de Broglie wave.
For the other two modes, the dispersion relation is instead given by
\begin{equation}
  \label{eq:c2k}
   \omega_2^2\approx\frac{E_{22}D_{11}}{B^2+A_{22}D_{11}}\,\frac{k^2}{a^2}\, .
\end{equation}
So, we have two propagation speeds, defined as the group velocity $a\partial\omega/\partial k$, whose squared expressions are given by
\begin{align}
  c_1^2&={\left( a \frac{\partial \omega_1}{\partial k}\right)}^2 \approx -{\frac {256}{3}}\,{\frac { \dot H
 ^{2} \Delta }{ \left( 8 \,F_{\GB} H ^{2}+16\, H ^{4}\xi_{\GB}+F_R \right) ( F+4\,H\,\dot\xi ) }}\ \frac{k^2}{a^2},\label{c1} \\
  c_2^2&={\left( a \frac{\partial \omega_2}{\partial k}\right)}^2 \approx \frac{\dot p}{\dot \rho}=\left(\frac{\partial p}{\partial \rho}\right)_s\,.
\end{align} The Eq.\ (\ref{c1}) is obviously model dependent and can be used to constrain gravity models. Here we want to point out that the first two modes have a $k$-dependent $c_1^2$, a feature that does not exist in GR and in modified gravity theories such as $f(R)$ and $R+f(\GB)$. This feature only appears in the general modified gravity theories $f(R,\GB)$ with $\Delta \neq 0$, and affects only the scalar perturbations. The dispersion relation for those modes has been derived for the vacuum case in \cite{DeFelice09}, and its expression exactly coincides with Eq.\ (\ref{c1}). The only difference is the change in the background expansion of the universe due to the existence of a perfect fluid. This result was expected because a minimal coupling of matter to gravity should not affect the high-$k$ regime. As discussed in detail in \cite{DeFelice09}, Eq.~(\ref{c1}) shows that, for short wavelength modes, either the propagations become super-luminal if $c_1^2$ is positive or the spacetime becomes very unstable if $c_1^2$ is negative. Here we have explicitly confirmed that the inclusion of a perfect fluid does not alter the nature of this physical property at all, i.e.~we still have either a super-luminal propagation or a spacetime instability if $\Delta \ne 0$. Among these two characteristics, the second property might be inconsistent with a viable cosmology  and, for this reason, could be used to rule out models with $\Delta \ne 0$. The first property, the superluminal propagation, is not a problem per se as it does not directly violate causality on the cosmological FLRW background. However it indicates that the UV completion of the theory may not be Lorentz invariant \cite{DGPSL}. We conclude that $f(R,\GB)$ modified gravity theories can indeed be classified according to some physical properties: if $\Delta \neq 0$, one propagation speed is scale-dependent ($c_1 \propto k^1$); if $\Delta =0$, the propagation speeds are all scale-independent. So direct observations might disentangle them.

The other two modes having a squared speed of propagation $c_2^2$ represent the propagation of scalar perturbations arising from the perfect fluid. Because of its minimal coupling, the propagation nature of a perfect fluid is not affected by the modification of gravity. The remaining two modes only change on cosmological time scale, i.e.\ $\omega^2 \propto k^0$, as shown in the appendix \ref{Kappo}.

Let us now consider the subset of theories for which the condition $\Delta =0$ is fulfilled. It is obvious that both $f(R)$ and $R+f(\GB)$ theories belong to this class. As shown in \cite{DeFelice09}, however, there is an infinite number of other $f(R,\GB)$ theories that belong to this class. If $\Delta=0$, then the $B$ and $D_{11}$ terms vanish in Eq.\ (\ref{appeqforVa}). Thus we expect that the ultra-violet behavior drastically changes, and the de Broglie type dispersion relation should be modified. To see this explicitly, let us solve the resulting evolution equations for large $k$'s:
\begin{equation}
A_{ab} {\ddot V_b}+E_{ab} \frac{k^2}{a^2}V_b=0. \label{appeqforVasp}
\end{equation}
We find that the modes propagate with speeds given now by
\begin{align}
  c_1^2&\approx-\frac13\,{\frac {16\, \dot\xi  H ^{2}\ddot\xi-64\, H ^{3} \dot\xi ^{2}-64\, \dot\xi ^{2} \dot H H  -12\, \dot\xi  H ^{2}F -16\, \dot F H \dot\xi -16\, \dot\xi \dot H F +4\, \ddot\xi \dot F -3\,F \dot F }{16\, H ^{3} \dot\xi ^{2}+4\, \dot\xi  H ^{2}F +4\, \dot F H \dot\xi +F \dot F }}\,,\\
  c_2^2&\approx \frac{\dot p}{\dot \rho}\, . 
\end{align}
In the case of $f(R)$ theories, $\xi=0$ and we recover the well-known result $c_1^2=1$. 
Theories which are supposed to lead to dark energy at late times need to reduce to GR at early times. 
In this case one typically has $f_{\GB \GB}H^6\ll1$ and $f_{RR}H^2\ll1$ \cite{fRviable,GaussBN}, so that $\dot F$ and $\dot \xi$ can be considered as corrections with respect to other GR-like leading terms.
In this limit, the expression for $c_1^2$ becomes
\begin{equation}
  \label{eq:gbnt}
  c_1^2\approx1+\frac{16}3\,\frac{\dot H\dot\xi}{\dot F+4H^2\dot\xi}\, .
\end{equation}
When this happens, also the background is GR-dominated and $2\dot H\approx-3(1+p/\rho)H^2$, so that
\begin{equation}
  \label{eq:c1spcaprx}
  c_1^2\approx1-2\left(1+\frac p \rho\right)\frac1{1+\frac{\dot F}{4H^2\dot\xi}}\, .
\end{equation}
If $\dot F\ll4H^2\dot\xi$, we are in the Gauss-Bonnet gravity regime, for which evidently the squared speed of propagation becomes in general negative. This result, applied to the $R+f(\GB)$ theory, confirms the one found in \cite{TsujiInst}. If $\dot F\gg4H^2\dot\xi$, then we are in the $f(R)$ regime for which $c_1^2\approx 1$. 
Therefore, one has to require these theories to be closer to the $f(R)$ regime rather than to the Gauss-Bonnet one.

\subsection{Ghost conditions}

Once we imposed the speed of propagation to be positive, one should also make sure that the modes are not ghost-like. This is achieved by demanding the matrix $A$ to be positive definite. For that purpose, one starts with a simple field redefinition
\begin{align}
  V_1&={\cal C}_1\,W_1\,,\\
  V_2&=-\frac{A_{12}}{A_{22}}\,{\cal C}_1\,W_1+{\cal C}_2\,W_2\, .
\end{align}
In this case, the kinetic matrix $\tilde A$ for the fields $W_{1,2}$ diagonalizes into
\begin{equation}
  \label{eq:diagW}
 \tilde A={\rm diag} ({\cal C}_1^2\,A_{22}^{-1}\,\det A ,\;{\cal C}_2^2\, A_{22}),    
\end{equation}
and the additional conditions to impose are
\begin{align}
A_{22}&=\frac{n^2}{\rho+p}\,\frac{\dot\rho}{\dot p}+
\frac{9216\pi^2 G^2 n^2 H^4\dot H^2 (p+\rho)^2\xi_{\GB}\Delta}{\dot p\,J}\geq0\,,\\
  \det A&=-\frac{9}{4J}\,H\,n^2\,(\dot F+4H^2\dot\xi)^2(F_{\GB}+4H^2\xi_{\GB})^2(F+4H\dot\xi)
\left[1+\frac{\Delta}{(F_{\GB}+4H^2\xi_{\GB})^2}\right]\geq0\, ,
\end{align}
where $J$ is defined in appendix \ref{coeffo}. There is an extra no-ghost condition
\begin{equation}
  \label{eq:Knog}
  K\geq0, 
\end{equation}
that can be derived once the field $\delta s$ is also integrated out (see appendix \ref{Kappo}). 
It amounts to impose a negative coefficient for $\delta s^2$ in the action (\ref{secondaction}).

\section{Vector modes}\label{sec:vector}

The vector modes treated from an action point of view were discussed for General Relativity in \cite{DeGeSu}. Along the same lines, we find here the action for the vector modes in $f(R,\GB)$ theories. As for the metric, we introduce two vector fields such that $g_{0i}=-a\,G_i$ and $g_{ij}=a^2(C_{i,j}+C_{j,i})$. We define then the divergence part of the perturbed 3-velocity, $\delta u_i$ of the perfect fluid as $u^V_i$. In terms of the gauge invariant $V_i=-(G_i+aC_i)$, and fixing the gauge $\delta B=0$,  we find the following action
\begin{equation}
  \label{eq:azzio}
  S=\tfrac12\int\de t\de x^3 a^3\left[
\frac{F+4H\dot\xi}{16\pi Ga^2}\,(\partial_jV_i)\,(\partial_j V_i)+2(\rho+p)\,u^V_i\,\dot C_i-\frac{\rho+p}{a^2}u^V_iu^V_i+2\frac{\rho+p}a\,u^V_i\,V_i
\right]\, .
\end{equation}
The variations of this action with respect to the fields $V_i$ and $C_i$ lead respectively to
\begin{align}
  \vec\nabla^2V_i&=\frac{16\pi G(\rho+p)}{F+4H\dot\xi}\,a\,u^V_i\,,\\
  \frac{\de}{\de t}[(\rho&+p)\,a^3\,u^V_i]=0\, .
\end{align}
This result implies that $u^V_i$ is a decaying mode, and so is $V_i$, in general.

Using a Fourier decomposition for all the fields, and integrating out both $u^V_i$ and $V_i$, one obtains the following action
\begin{equation}
  S=\tfrac12\,\int \de t\de k^3\,a^3\,Q\,\dot C_i\,\dot C_i\, ,
\end{equation}
where
\begin{equation}
  \label{eq:QV}
  Q=\frac{(\rho+p)k^2a^2\,(F+4H\dot\xi)}{(F+4H\dot\xi)k^2+16\pi G(\rho+p)a^2}\, .
\end{equation}
Consequently, in order not to have ghosts for any value of $k$, one needs to impose the condition
\begin{equation}
  \label{eq:VVgh}
  F+4H\dot\xi\geq0\, .
\end{equation}

\section{Tensor modes}\label{sec:tensor}

These modes are the easiest ones to study, as no contributions arise from the new scalars degrees of freedom and the  perfect fluid matter Lagrangian. In this case the action for the $H_{ij}$ modes, with $H^{ij}{}_{,j}=0=H^i{}_{i}$, can be written after decomposing them into the two polarization modes $\epsilon_{\otimes},\epsilon_{\oplus}$. Calling $A_{\otimes}$ and $A_{\oplus}$ their corresponding amplitudes, one finds
\begin{equation}
  \label{eq:GW}
  S=\tfrac12\int\de t\,\de x^3\,a^3\left[\frac{F+4H\dot \xi}{16\pi G}\,\dot A_{\otimes}^2-\frac{F+4\ddot\xi}{16\pi G a^2}(\vec\nabla A_{\otimes})^2 \right] ,
\end{equation}
and a similar action for $A_{\oplus}$. Therefore the no-ghost condition coincides with the one given by the vector modes. However, these modes do propagate and their speed of propagation reduces to
\begin{equation}
  \label{eq:GWc2}
  c^2_{\otimes}=\frac{F+4\ddot\xi}{F+4H\dot\xi}\geq0\, .
\end{equation}
This is an independent condition, which implies the classical stability of the spin-2 metric perturbations.

\section{conclusions}\label{sec:concl}

We have developed the linear cosmological perturbations for $f(R,\GB)$ theories where a perfect fluid
is minimally coupled to gravity.

For the scalar type perturbation, we found that there are in general six degrees of freedom. Two modes out of six, which do not exist in GR and appear only in modified gravity theories, represent the gravity perturbations. If $f_{R R} f_{\GB \GB} \neq f_{R \GB}^2$, these modes obey a dispersion relation which is the same as the one for the non-relativistic de Broglie wave: $\omega^2=L(t) k^4$. Correspondingly, the perturbations either propagate with super-luminal speeds if $L(t)$ is positive or are highly unstable if $L(t)$ is negative on small scales. The apparent expression of $L(t)$ exactly coincides with the vacuum case given in \cite{DeFelice09}. This result is sensible as a minimally coupled perfect fluid cannot affect the nature of the de Broglie type dispersion relation which solely comes from the modification of gravity. The other two modes represent density and velocity perturbations of a perfect fluid, which propagate with a sound velocity $c_s^2={\dot p}/{\dot \rho}$, just as in GR. The remaining two modes represent the entropy and temperature perturbations of a perfect fluid. These modes completely decouple from the other modes for a perfect fluid with a barotropic equation of state.

For the vector type perturbation, we found that the perturbations decay in time, which is the same as in GR.
For the tensor type perturbation, the perturbations can be either super-luminal or sub-luminal, depending on the model. We found that both perturbations must fulfill the same no-ghost condition.

The advantage of the action approach presented here is that we can determine the sign of the kinetic terms for the perturbation variables. The kinetic terms must be positive to avoid ghost, which put another constraint on $f(R,\GB)$ theories. We found all the no-ghost conditions for the modes. These conditions must be satisfied as well as the stability conditions for the perturbation squared speeds in order to construct viable $f(R,\GB)$ theories.

\begin{acknowledgments}
The work of A.D.F.\ was supported by the Grant-in-Aid for Scientific Research Fund of the JSPS No.\ 09314.
This work is supported by the Belgian Federal Office for Scientific, Technical and Cultural Affairs through the Interuniversity Attraction Pole P6/11. 
T.S.\ is supported by a Grant-in-Aid for JSPS Fellows.
\end{acknowledgments}

\appendix

\section{Coefficients in the action (\ref{secondaction})}
\label{coeffo}

If we define
\begin{align}
  \Gamma_1&=F+4H\dot\xi\, \\
  \Gamma_2&=\dot F+12H^2\dot\xi+2FH\, \\
\Gamma_3&=\dot p\,(\dot F+4H^2\dot\xi)+8\pi G(p+\rho)^2\, \\
\Gamma_4&=F_{\GB}+4H^2\xi_{\GB}\,,\\
\Gamma_5&=(\rho+p)\,\dot T-\dot p T\,,\\
\Gamma_7&=\dot F+4H^2\dot\xi\, .
\end{align}
and 
\begin{align}
  J&=16\pi G\,\Gamma_2^2 \,\dot p\,H^4 \xi_{\GB} ^{2}+ 8\pi G\Gamma_2^2\, H^2\,\dot p\,F_{\GB} \xi_{\GB}+ \pi G\,\Gamma_2^2\,\dot p \,(\Delta+F_{\GB}^2)\notag\\
&+ 768\pi G H^3\dot H^2[4\pi G (\rho+p) ^{2}-H\dot p\Gamma_1]\,\xi_{\GB}\,\Delta,
\end{align}
the elements for the kinetic matrix $A$ are
\begin{align}
  A_{11}&=\frac{3\Gamma_1}{4\pi G\Gamma_2^2}
  \left[\Gamma_6^2-\frac{16\pi G H\Gamma_1 (\rho+p)^2}{\dot p}\right]
+\frac{576H^4\dot H^2\Gamma_1^2\Gamma_3^2\xi_{\GB}\Delta}{\dot p\,\Gamma_2^2\,J}\,\\
A_{12}&=-\frac{6nH(p+\rho)\Gamma_1}{\dot p\,\Gamma_2}
+\frac{2304\pi G n (\rho+p) H^4\dot H^2\Gamma_1\Gamma_3\xi_{\GB}\Delta}{\dot p\,\Gamma_2\,J}\,\\
A_{22}&=\frac{n^2}{\rho+p}\,\frac{\dot\rho}{\dot p}+
\frac{9216\pi^2 G^2 n^2 H^4\dot H^2 (p+\rho)^2\xi_{\GB}\Delta}{\dot p\,J}\, .
\end{align}
and
\begin{align}
  B&=-\frac{384\pi G n H^2\dot H^2(\rho+p)\Gamma_6\xi_{\GB}\Delta}J\,,\\
 \label{eq:D11}
  D_{11}&=-\frac{16\Gamma_6^2\xi_{\GB} \dot H ^{2}\dot p\Delta}J\, ,\\
  C&=\frac{12nH\Gamma_1}{\Gamma_2^2}\left[\frac{4\pi G(p+\rho)^2}{\dot p}+H\Gamma_1+\Gamma_6\right]
  -\frac{2304 \pi G n H^4\dot H^2 \Gamma_1\Gamma_3^2\xi_{\GB}\Delta}{\dot p\,\Gamma_2^2\,J}\, .
\end{align}
The interaction coefficients between the two propagating fields $\phi$ and $v$ with $\delta s$ are the following ones
\begin{align}
E_{13}&=\frac{128\pi G\Gamma_5nH\dot H^2 \Gamma_6\,\xi_{\GB}\Delta}{J}\,,\\
C_{13}&=-\frac{4n\Gamma_1\Gamma_5}{\dot p\Gamma_2}+\frac{1536\pi G n\Gamma_5\Gamma_1\Gamma_3H^3\dot H^2\xi_{\GB}\Delta}{J\dot p\,\Gamma_2}\,,\label{c13}\\
C_{23}&=-\frac{2n^2\dot T}{\dot p}+\frac{6144\pi^2G^2 n^2(\rho+p)\dot H^2 H^3\Gamma_5\xi_{\GB}\Delta}{\dot p\, J}\,,\label{c23}\\
m_{23}&=\frac{8\pi Gn^2\Gamma_5}{\dot p\,\Gamma_2}
-\frac{3072\pi^2G^2H^3\dot H^2 n^2\Gamma_3\Gamma_5 \xi_{\GB}\Delta}{\dot p\,J\,\Gamma_2}\,. \label{m23}
\end{align}
The coefficient for $\delta s^2$ is given by
\begin{align}
  K&=nT {\left( \frac{\partial T}{\partial \mu} \right)}_s+ n\left(\frac{\partial T}{\partial s}\right)_{\!\mu}-\frac{1024\pi^2G^2H^2\dot H^2n^2\Gamma_5^2\xi_{\GB}\Delta}{\dot p\,J}\,.
\end{align}
For the Laplacian matrices $E$ and $M$, it is simple to write down the elements
\begin{equation}
E_{22}=\frac{n^2}{p+\rho},~~M_{11}^2=0.
\end{equation}
The other elements (i.e., $E_{11}$, $E_{12}$, $M^2_{12}$, and $M^2_{22}$), because of their complexity, cannot fit into a standard paper.
We provide a file to download them \cite{fileup}.

\section{Useful relations for perfect fluids on a FLRW background}\label{fluido}

For a perfect fluid we have the first principle of thermodynamics---together with two equations of state $n=n(\mu,s)$ and $T=T(\mu,s)$---written as
\begin{equation}
  \label{eq:apf1}
  \de p=n\de \mu-nT\de s\, .
\end{equation}
Consequently, if $p=p(\mu,s)$ we have the following useful relations:
\begin{equation}
\left(\frac{\partial p}{\partial \mu}\right)_{\!s}=n, ~~~~~~\left(\frac{\partial p}{\partial s}\right)_{\!\mu}=-nT \label{thermal}
\end{equation}
and, by analyticity,
\begin{equation}
\left(\frac{\partial n}{\partial s}\right)_{\!\mu}=-\left(\frac{\partial nT}{\partial \mu}\right)_{\!s}. \label{maxwell}
\end{equation}

In a FLRW universe, $\dot s=0$ implies that
\begin{equation}
  \label{eq:apf2}
  \dot\mu=\dot p/n\, .
\end{equation}
By differentiating the identity
\begin{equation}
\rho \equiv \mu n-p, \label{mudef}
\end{equation}
and using Eq.\ (\ref{eq:apf2}) together with the conservation of the particle number ($n=N/a^3$), one then recovers the continuity equation
\begin{equation}
  \dot\rho=-3H(\rho+p)\, .
\end{equation}
These relations are quite standard. Let us consider now four limiting cases in $f(R,\GB)$ theories.

\vspace{5mm}

\begin{center}
{\bf Case I: $\Gamma_5=0$}
\end{center}

The combination $\Gamma_5=(\rho+p){\dot T}-{\dot p}T$ appears many times in the coefficients of the action (\ref{secondaction}).
In particular, $\delta s$ is no longer coupled to $V_1$ when $\Gamma_5$ vanishes.
So, it is instructive to understand for which perfect fluids it vanishes, i.e., 
\begin{equation}
\label{eq:tmu}
(p+\rho)\left(\frac{\partial T}{\partial\mu}\right)_{\!s}\dot\mu=T\dot p\, ,
\end{equation}
since $T=T(\mu,s)$ with ${\dot s}=0$ on the FLRW background.
The Eqs.\ (\ref{eq:apf2}) and (\ref{mudef}) imply then that
\begin{equation}
\mu\left(\frac{\partial T}{\partial\mu}\right)_{\!s}=T\, ,
\end{equation}
namely
\begin{equation}
\label{eq:baroTs}
T=f(s)\,\mu\, .
\end{equation}
For a system with two thermodynamical degrees of freedom, this condition is equivalent to having a barotropic equation of state. Indeed, if
$\mu=\mu (\rho,s)$ and $\rho=\rho(\mu,s)$, we have
\begin{equation}
\left(\frac{\partial\mu}{\partial s}\right)_{\!\rho} =-\left(\frac{\partial\mu}{\partial \rho}\right)_{\!s} \left(\frac{\partial \rho}{\partial s}\right)_{\!\mu}. \label{b10}
\end{equation}
From Eqs. (\ref{b10}), (\ref{mudef}), (\ref{thermal}) and (\ref{maxwell}) we respectively infer 
\begin{align}
\left(\frac{\partial\mu}{\partial s}\right)_{\!\rho} &=-\frac{\left(\frac{\partial \rho}{\partial s}\right)_{\!\mu}}%
{\left(\frac{\partial\rho}{\partial\mu}\right)_{\!s}}
=-\frac{\mu \left(\frac{\partial n}{\partial s}\right)_{\!\mu}-\left(\frac{\partial p}{\partial s}\right)_{\!\mu}}{n+\mu \left(\frac{\partial n}{\partial \mu}\right)_{\!s}-\left(\frac{\partial p}{\partial \mu}\right)_{\!s}}
=-\frac{\mu \left(\frac{\partial n}{\partial s}\right)_{\!\mu}+nT}{\mu \left(\frac{\partial n}{\partial \mu}\right)_{\!s}}=-\frac{\mu \bigg[ -\left(\frac{\partial n}{\partial \mu}\right)_{\!s} T-n \left(\frac{\partial T}{\partial \mu}\right)_{\!s} \bigg]+nT}{\mu \left(\frac{\partial n}{\partial \mu}\right)_{\!s}}.
\end{align}
Using now Eq.~(\ref{eq:baroTs}), we eventually find
\begin{equation}
\left(\frac{\partial\mu}{\partial s}\right)_{\!\rho}=-\frac{\mu \left[-\left(\frac{\partial n}{\partial \mu}\right)_{\!s} T-n T/\mu\right]+nT}{\mu \left(\frac{\partial n}{\partial \mu}\right)_{\!s}}=T\,,
\end{equation}
so that $(\partial p/\partial s)_\rho=0$, or, equivalently, $p=p(\rho)$. 
This last statement is a consequence of Eq.\ (\ref{eq:apf1}), when we change the two independent variables from $(\mu,s)$ to $(\rho,s)$. After doing this, one should consider $\mu=\mu(\rho,s)$, and the first law of thermodynamics implies $(\partial p/\partial s)_\rho=n[(\partial\mu/\partial s)_\rho-T]$.

From Eq.~(\ref{c23}), we obtain $C_{23}=-2n f(s)$ such that the conservation of the particle number $N=a^3n$ together with the conservation of the entropy per particle $s$ imply $\partial_t (a^3 C_{23})=0$. Consequently, the field $\delta s$ decouples from both $V_1$ and $V_2$ in the action (\ref{secondaction}). Indeed, $C_{13}$ defined in Eq.~(\ref{c13}) and $m_{23}$ defined in Eq.~(\ref{m23}) also vanish if $\Gamma_5=0$.

\vspace{5mm}

\begin{center}
{\bf Case II: $K=0$}
\end{center}

It is also useful to understand for which equations of state the coefficient $K$ for $\delta s^2$ vanishes, at least in the $\Delta=0$ consistent case. This happens for
\begin{equation}
T \left(\frac{\partial T}{\partial\mu}\right)_{\!s}+\left(\frac{\partial T}{\partial s}\right)_{\!\mu}=0,
\end{equation}
or, equivalently,
\begin{equation}
T=-\frac{\left(\frac{\partial T}{\partial s}\right)_{\!\mu}}{\left(\frac{\partial T}{\partial\mu}\right)_{\!s}}=
\left(\frac{\partial\mu}{\partial s}\right)_{\! T}\, 
\end{equation}
if, again, the system has two degrees of freedom, i.e., if $T=T(\mu,s)$. The solution
\begin{equation}
\mu=\mu(T,s)=Ts+g(T)\, 
\end{equation}
implies that
\begin{equation}
\de p=n\de\mu-nT\de s=n[s+g'(T)]\,\de T\, .
\end{equation}
In order that $\de p$ be an exact differential we need then to impose that
\begin{equation}
n=n(T,s)=\frac{h(T)}{s+g'(T)}\, ,
\end{equation}
so that we finally find $(\partial p/\partial s)_T=0$, or, equivalently, $p=p(T)$. 

\vspace{5mm}

\begin{center}
{\bf Case III: $K=0,\Gamma_5=0$}
\end{center}

There is an equation of state for which both $K$ and $\Gamma_5$ vanish, even when $\Delta\neq0$. 
This case corresponds to $f(s)=\frac{1}{s+c}$ and $g(T)=c T$ ($c$ is a constant). 
In this case we find that
\begin{equation}
T=\frac{\mu}{s+c}\, ,
\end{equation}
and $\rho=\rho(T)$ since $p=p(\rho)$ and $p=p(T)$. A radiation fluid fulfills these properties.

\vspace{5mm}

\begin{center}
{\bf Case IV: ideal gas with $\Delta=0$}
\end{center}

It is also interesting to derive $K$ and $\Gamma_5$ for a classical monoatomic ideal gas.
The thermodynamical quantities for this case are given by
\begin{align}
n(\mu,s)&=\frac{m^{3/2}}{5 \sqrt{5} \pi^{3/2}}\,(\mu-m)^{3/2}\,\exp\!\left(\tfrac52-s\right),\\
T(\mu,s)&=\tfrac{2}{5} (\mu-m)\,,\\
p(\mu,s)&=n\, T, \\
\rho (\mu,s)&=m\,n+\tfrac32\,n\,T\,,
\end{align}
where $m$ is the mass of an atom. We have used the Sackur--Tetrode equation for a monoatomic atom. 
The energy density $\rho$ is calculated from the identity $\mu n=\rho+p$.
Assuming a model satisfying $\Delta=0$, we find that $K$ and $\Gamma_5$ are written as
\begin{equation}
K=\tfrac{2}{5}\, n\,T,  \qquad\Gamma_5=-2m\,H\,p\, .   
\end{equation}
Since $T>0$, we have $K>0$, so that no ghost appears for an ideal gas, as expected from our real experience.

\section{Eliminating the auxiliary field}\label{Kappo}
Let us integrate out $\delta s$ from the action. In that case, $\delta \theta_M$ gets a kinetic term quadratic in its time derivative. It is then useful to define a new variable
\begin{equation}
V_3 \equiv \delta \theta_M-\frac{C_{a3}}{2n}\,V_a\, ,
\end{equation}
which can be considered as a redefinition for $\delta\theta_M$. Keeping only the relevant higher order terms in $k$, the action (\ref{secondaction}) becomes
\begin{equation}
S=\tfrac12\int \de t\, \de^3x\, a^3 \left( A_{ab} {\dot V_a} {\dot V_b}-a^{-2}F_{ab} {\vec \nabla} {V_a} \cdot {\vec \nabla}\dot V_b-a^{-4}D_{ab} \vec\nabla^2 V_a \vec\nabla^2 V_b -a^{-2} L_{ab} {\vec \nabla} V_a \cdot {\vec \nabla} V_b \right), \label{integralout}
\end{equation}
where now both $a$ and $b$ run from 1 to 3, and $A_{ab},F_{ab},D_{ab}$ and $L_{ab}$ are all 3 $\times$ 3 matrices given by
\begin{align}
&A=\left( \begin{array}{rrr}
A_{11} & A_{12} & 0 \\
A_{12} & A_{22} & 0 \\
0 & 0 & A_{33} \\
\end{array} \right), \hspace{3mm}
F=\left( \begin{array}{rrr}
0 & B & F \\
-B & 0 & 0 \\
-F & 0 & 0 \\
\end{array} \right), \hspace{3mm}
D=\left( \begin{array}{rrr}
Q & 0 & 0 \\
0 & 0 & 0 \\
0 & 0 & 0 \\
\end{array} \right), \\
&L=\left( \begin{array}{ccc}
E_{11}+\frac{E_{13}}{2K} {(\dot C_{13}+3HC_{13})} & E_{12}+\frac{E_{13}}{4K} ({\dot C_{23}+3HC_{23}-2m_{23}}) & 
\frac{n}{4K^2}[E_{13}(\dot K+2HK)-K\dot E_{13}] \\
E_{12}+\frac{E_{13}}{4K} ({\dot C_{23}+3HC_{23}-2m_{23}}) & E_{22} & 0 \\
\frac{n}{4K^2}[E_{13}(\dot K+2HK)-K\dot E_{13}] & 0 & 0 \\
\end{array} \right).
\end{align}
Here 
\begin{equation}
A_{33}=\frac{n^2}{K},\qquad F=\frac{nE_{13}}{2K},\qquad Q=D_{11}-\frac{E_{13}^2}{4K}, \label{eq1}
\end{equation} 
and all the other quantities have already been defined in the appendix A.

The corresponding dispersion relation is obtained from an eigenvalue equation
\begin{equation}
\det \left( -\omega^2 A+i \omega \frac{k^2}{a^2} F+\frac{k^4}{a^4} D +\frac{k^2}{a^2} L \right)=0.
\end{equation}
This is 6-th order algebraic equation for $\omega$. Using the above equations, we confirm that two solutions coincide with $\omega_1$ given by Eq.(\ref{eq:cs2k2}) and the other two solutions coincide with $\omega_2$ given by Eq.(\ref{eq:c2k}). We also check that the remaining two solutions are independent of $k$, which means that these modes evolve only on the cosmic time scale.

The action (\ref{integralout}) leads to the extra no-ghost condition $A_{33}\geq0$, which implies $K\geq0$. Although classically the equation of motion for $V_3$ depends on the evolution of $V_1$ and $V_2$, its kinetic term should be positive in order to have a sensible spectrum for the modes at the quantum level.



\begin{thebibliography}{10}

\bibitem{review}
  A.~De Felice and S.~Tsujikawa,
  arXiv:1002.4928 [gr-qc].

\bibitem{Starob}
  A.~A.~Starobinsky,
  Phys.\ Lett.\  B {\bf 91}, 99 (1980).

\bibitem{fRori}
S.~Capozziello,
Int.\ J.\ Mod.\ Phys.\  D {\bf 11}, 483 (2002);
S.~Capozziello, S.~Carloni and A.~Troisi,
Recent Res.\ Dev.\ Astron.\ Astrophys.\  {\bf 1}, 625 (2003);
S.~Capozziello, V.~F.~Cardone, S.~Carloni and A.~Troisi,
Int.\ J.\ Mod.\ Phys.\  D {\bf 12}, 1969 (2003);
S.~M.~Carroll, V.~Duvvuri, M.~Trodden and M.~S.~Turner,
Phys.\ Rev.\  D {\bf 70}, 043528 (2004).

\bibitem{fRviable}
L.~Amendola, R.~Gannouji, D.~Polarski and S.~Tsujikawa,
Phys.\ Rev.\  D {\bf 75}, 083504 (2007);
B.~Li and J.~D.~Barrow,
Phys.\ Rev.\  D {\bf 75}, 084010 (2007);
W.~Hu and I.~Sawicki,
Phys.\ Rev.\  D {\bf 76}, 064004 (2007);
A.~A.~Starobinsky,
JETP Lett.\  {\bf 86}, 157 (2007);
S.~A.~Appleby and R.~A.~Battye,
Phys.\ Lett.\  B {\bf 654}, 7 (2007); 
S.~Tsujikawa,
Phys.\ Rev.\  D {\bf 77}, 023507 (2008);
E.~V.~Linder,
arXiv:0905.2962 [astro-ph.CO];
L.~Amendola and S.~Tsujikawa,
Phys.\ Lett.\  B {\bf 660}, 125 (2008).

\bibitem{fRmatter}
S.~M.~Carroll, I.~Sawicki, A.~Silvestri and M.~Trodden,
New J.\ Phys.\  \textbf{8}, 323 (2006);
Y.~S.~Song, W.~Hu and I.~Sawicki,
Phys.\ Rev.\  D {\bf 75}, 044004 (2007);
I.~Sawicki and W.~Hu,
Phys.\ Rev.\  D {\bf 75}, 127502 (2007);
R.~Bean, D.~Bernat, L.~Pogosian, A.~Silvestri and M.~Trodden,
Phys.\ Rev.\  D {\bf 75}, 064020 (2007);
T.~Faulkner, M.~Tegmark, E.~F.~Bunn and Y.~Mao,
Phys.\ Rev.\  D {\bf 76}, 063505 (2007);
L.~Pogosian and A.~Silvestri,
Phys.\ Rev.\  D {\bf 77}, 023503 (2008).

\bibitem{PQR}
S.~M.~Carroll, A.~De Felice, V.~Duvvuri, D.~A.~Easson, M.~Trodden and M.~S.~Turner,
Phys.\ Rev.\  D {\bf 71}, 063513 (2005);
G.~Calcagni, S.~Tsujikawa and M.~Sami,
Class.\ Quant.\ Grav.\  {\bf 22}, 3977 (2005).

\bibitem{Hanlon}
J.~O' Hanlon,
Phys.\ Rev.\ Lett.\  {\bf 29}, 137 (1972);
P.~Teyssandier and P.~Tourrenc,
J.\ Math.\ Phys.\  {\bf 24}, 2793 (1983);
T.~Chiba,
Phys.\ Lett.\  B {\bf 575}, 1 (2003).

\bibitem{GaussBN}
S.~Nojiri, S.~D.~Odintsov and M.~Sasaki,
Phys.\ Rev.\  D {\bf 71}, 123509 (2005);
T.~Koivisto and D.~F.~Mota,
Phys.\ Lett.\  B {\bf 644}, 104 (2007);
T.~Koivisto and D.~F.~Mota,
Phys.\ Rev.\  D {\bf 75}, 023518 (2007);
S.~Kawai, M.~a.~Sakagami and J.~Soda,
Phys.\ Lett.\  B {\bf 437}, 284 (1998);
S.~Nojiri and S.~D.~Odintsov,
Phys.\ Lett.\  B {\bf 631}, 1 (2005);
A.~De Felice and S.~Tsujikawa,
Phys.\ Lett.\  B {\bf 675}, 1 (2009);
T.~P.~Sotiriou,
arXiv:0710.4438 [gr-qc];
K.~Uddin, J.~E.~Lidsey and R.~Tavakol,
Gen.\ Rel.\ Grav.\  {\bf 41}, 2725 (2009);
G.~Cognola, E.~Elizalde, S.~Nojiri, S.~D.~Odintsov and S.~Zerbini,
Phys.\ Rev.\  D {\bf 73}, 084007 (2006);
B.~Li, J.~D.~Barrow and D.~F.~Mota,
Phys.\ Rev.\  D {\bf 76}, 044027 (2007);
S.~Y.~Zhou, E.~J.~Copeland and P.~M.~Saffin,
JCAP {\bf 0907}, 009 (2009);
A.~De Felice and S.~Tsujikawa,
Phys.\ Rev.\  D {\bf 80}, 063516 (2009).

\bibitem{TsujiInst}
  A.~De Felice, D.~F.~Mota and S.~Tsujikawa,
  Phys.\ Rev.\  D {\bf 81}, 023532 (2010)
  [arXiv:0911.1811 [gr-qc]].
  
\bibitem{DeFelice09}
A.~De Felice and T.~Suyama,
JCAP {\bf 0906}, 034 (2009).

\bibitem{DeSuScal}
  A.~De Felice and T.~Suyama,
  Phys.\ Rev.\  D {\bf 80}, 083523 (2009)
  [arXiv:0907.5378 [astro-ph.CO]].

\bibitem{cosmoreview}
J.~M.~Bardeen,
Phys.\ Rev.\  D {\bf 22}, 1882 (1980);
H.~Kodama and M.~Sasaki,
Prog.\ Theor.\ Phys.\ Suppl.\  {\bf 78} (1984) 1;
V.~F.~Mukhanov, H.~A.~Feldman and R.~H.~Brandenberger,
Phys.\ Rept.\  {\bf 215}, 203 (1992).

\bibitem{ghost1}
K.~S.~Stelle,
Gen.\ Rel.\ Grav.\  {\bf 9}, 353 (1978);
N.~H.~Barth and S.~M.~Christensen,
Phys.\ Rev.\ D {\bf 28}, 1876 (1983);
N.~Boulanger, T.~Damour, L.~Gualtieri and M.~Henneaux,
  Nucl.\ Phys.\  B {\bf 597}, 127 (2001)
  [arXiv:hep-th/0007220];
A.~De Felice, M.~Hindmarsh and M.~Trodden,
JCAP {\bf 0608}, 005 (2006);
G.~Calcagni, B.~de Carlos and A.~De Felice,
Nucl.\ Phys.\ B {\bf 752}, 404 (2006).

\bibitem{Schutz}
  B.~F.~Schutz,
  Phys.\ Rev.\  D {\bf 2}, 2762 (1970);
  B.~F.~Schutz and R.~Sorkin,
  Annals Phys.\  {\bf 107}, 1 (1977).

\bibitem{DeGeSu}
  A.~De Felice, J.~M.~Gerard and T.~Suyama,
  Phys.\ Rev.\  D {\bf 81}, 063527 (2010)
  [arXiv:0908.3439 [gr-qc]].

\bibitem{DGPSL}
  K.~Hinterbichler, A.~Nicolis and M.~Porrati,
  JHEP {\bf 0909}, 089 (2009)
  [arXiv:0905.2359 [hep-th]].

\bibitem{fileup}
{\tt http://sites.google.com/site/adefelic/ricerca/extraterms}

\end{thebibliography}
\end{document}